\documentclass[onecolumn,showpacs,manuscript]{revtex4}
\usepackage{dcolumn}
\usepackage{amsmath}
\usepackage{amssymb}
\usepackage[dvips]{epsfig}  

\makeatletter
\def\btt#1{\texttt{\@backslashchar#1}}%
\DeclareRobustCommand\bblash{\btt{\@backslashchar}}%
\makeatother

\begin{document}

\title{New potential super-incompressible phase of ReN$_{2}$}

\author{Yanling Li$^{1,2,3}$, Zhi Zeng$^{1}$}
\affiliation {1.Key Laboratory of Materials Physics, Institute of
Solid State Physics, Chinese Academy of Sciences, Hefei
230031,People's Republic of China \\
2.Department of physics, Xuzhou Normal University, Xuzhou 221116,
People's Republic of China\\
3.Graduate School of the
Chinese Academy of Sciences, Beijing 100049, People's Republic of
China}
\date{\today}

\begin{abstract}
The structural, elastic, and electronic properties of
ReN$_{2}$ are investigated by first-principles calculations with density
functional theory. The obtained orthorhombic
$Pbcn$ structure is energetically the most stable structure at
ambient pressure.
ReN$_{2}$ is a metallic, superincompressible solid and presents a
rather elastic anisotropy. The estimated Debye temperature and hardness are 735 K and 17.1 GPa, respectively.
Its estimated hardness is comparative to that of Si$_{3}$N$_{4}$.

\end{abstract}



\maketitle

\section{Introduction}
Transition metal nitrides are of great technological and
fundamental importance because of their strength and durability as
well as their useful optical, electronic, and magnetic
properties.\cite{Crowhurst} By incorporating B, C, N, or O
atoms into the interstitial sites of transition metals with highly
valence electron density (VED), the compounds of transition metal with
highly hardness and strong incompressibility can be synthesized by
means of high pressure experiments.\cite {Kaner,Chung}
For these reasons, most recently, three transition metal
dinitrides, PtN$_{2}$,\cite {Crowhurst} IrN$_{2}$,\cite {Young} and
OsN$_{2}$,\cite {Young} have been successfully synthesized
under extreme conditions of pressure and temperature. Very large
bulk moduli observed in experiments make them be 
candidates of superhard materials, which could be used in
cutting tools and wear-resistant coatings. Meanwhile, many
theoretical investigations have been performed to explore their
structures, for that the crystal structure is an important
prerequisite of understanding their physical properties.
The search for the structural forms of PtN$_{2}$, OsN$_{2}$ and
IrN$_{2}$ with the lowest energy and mechanical stability usually
considers hypothetic structures of orthorhombic,
tetragonal, hexagonal and cubic crystal lattice. The structures of
PtN$_{2}$ and OsN$_{2}$ are indexed as cubic (pyrite structure,
$Pa\bar{3}$)\cite{Crowhurst,Young-1} and
orthorhombic (marcasite structure, $Pnnm$).\cite{Wu,Chen,Yu}
While it is very difficult to determine the structure of IrN$_{2}$
in experiment because of the large intensity ratio (1:100) of
X-ray diffraction peaks between nitride and pure metal. The
first-principles calculations have shown that IrN$_{2}$ is
monoclinic (CoSb$_{2}$ structure, $P2_{1}/c$).\cite{Yu} As a neighbor of
noble transition metal Os, the diboride of Re has been
successfully synthesized,\cite {Chung,Placa} however the
dinitride of Re to date has not been reported experimentally.
Theoretically, Zhao \emph{et al} suggested that ReN$_2$ with $P4_2/mnm$ structure is
thermodynamically stable at ambient conditions and up to 76 GPa.\cite{cms}
Here, we readdress structural and elastic properties of ReN$_2$ and find that a new
phase with $Pbcn$ symmetry is more stable than that with $P4_2/mnm$ symmetry.

\section{Computational details}
All calculations are performed by the CASTEP code\cite {Segall}
using \emph{ab initio} pseudopotentials based on DFT
with the exchange-correlation functional of Ceperley and Alder
as parameterized by Perdw and Zunger (LDA-CAPZ).\cite{LDA} All the
possible structures concerned are optimized by the BFGS
algorithm (proposed by Broyden, Fletcher, Goldfarb, and Shannon),\cite{BFGS}
which provides a fast way of finding the
lowest energy structure and supports cell optimization in the
CASTEP code. The interaction between the ions and the valence
electrons is described by using Vanderbilt's supersoft
pseudopotential\cite{Ultrasoft}.
We used a plane wave cutoff of 310 eV and a Brillouin zone sampling grid
spacing of 2$\pi$$\times$ 0.04 \AA$^{-1}$.
Pseudoatomic calculations are performed for N 2$s^{2}$2$p^{3}$ and
Re 5$s^{2}$5$p^{6}$5$d^{5}$6$s^{2}$.
In the geometrical optimization, all forces
on atoms are converged to less than 0.002 eV/\AA, all the stress
components are less than 0.02 GPa, and the tolerance in
self-consistent field (SCF) calculation is 5.0$\times$10$^{-7}$
eV/atom. Relaxation of the internal degrees of freedom is
performed at each unit cell compression or expansion.

The standard method is used to calculate the
elastic constants, in which the second
derivatives of the internal energy of a crystal are determined as a function of
properly chosen lattice distortions describing strains.
In practice, by applying small elastic strains, the elastic constants are obtained
from the change of energy or stress.
Elastic moduli are given by using the
elastic constants according to the Voigt-Reuss-Hill (VRH)
approximation.\cite{Hill} The Hill average, in general, is selected as the estimation of
bulk modulus and shear modulus.

\section{Results and discussions}
\subsection{Structural property}
To search for the most stable structure of ReN$_2$,
we consider monoclinic, orthorhombic, tetrahedral, hexagonal,
and cubic structures with a Re:N stoichiometry of 1:2, including these initial structures of
AuTe$_2$ ($C2/m$, No. 12), CoSb$_{2}$ ($P2_1/c$, No. 14), GeS$_2$ ($Fdd2$, No. 43), CaCl$_2$ ($Pnnm$, No. 58),
OsB$_2$ ($Pmmn$, No. 59), $\beta$-ReO$_2$ ($Pbcn$, No. 60), $Pbca$ (8$c$ Wycknoff sites, No. 61),
$Pnma$ (4$c$ Wycknoff sites, No. 62), ReSi$_2$ ($Immm$, No. 71), SiS$_2$ ($Ibam$, No. 72),
Cu$_2$Sb ($P4/nmm$, No. 129), rutile ($P4_2/mnm$, No. 136), MoSi$_2$ ($I4/mmm$, No. 139),
$P\bar{3}m1$ (1$a$ and 2$d$ Wycknoff sites, No 164),
CrSi$_2$ ($P6_222$, No. 180), Fe$_2$P ($P\bar{6}2m$, No. 189), AlB$_2$ ($P6/mmm$, No. 191),
ReB$_2$ ($P6_3/mmc$, No. 194), pyrite ($Pa\bar{3}$, No. 205),
fluorite ($Fm\bar{3}m$, No. 225), and NiTi$_2$ ($Fd\bar{3}m$, No. 227).
Combining total energy with elastic constants analysis,
four competing structures, i.e., $Pbcn$, $P4_2/mnm$, $Pmmn$, and $Fm\bar{3}m$,
satisfy the mechanical stability critetion,\cite{Beckstein,Born}
in which the latter three phases have been discussed by Zhao et al.\cite{cms}
The equilibrium volume \emph{V}$_{0}$ and bulk modulus B$_{0}$ are determined by
fitting the total energy as a function of volume to the 3rd-order
Birch Murnaghan equation of state (EOS).\cite{3rdEOS} Our
calculated equilibrium parameters (see Table I)
show that the rutile ($P4_2/mnm$) and the \emph{Pbcn} phases have
larger equilibrium volumes than the
fluorite (\emph{Fm$\bar{3}$m}) phase. From the calculated relative enthalpy $\Delta$H, one can see that the
\emph{Pbcn} phase is the most stable structure at zero pressure. In the optimized
\emph{Pbcn} structure, four rhenium atoms occupy 4\emph{c} Wycknoff
sites ($y$=0.1191) and eight N atoms hold 8\emph{d} sites
($x$=0.2663, $y$=0.3526, $z$=0.0871).
The nearest distance between Re (N) and N atoms is 1.955 (2.644) \AA.
The orthorhombic $Pbcn$ structure possesses a structure characterized by zigzag chains
of Re atoms propagating along the $c$ axis of the unit cell with
bond length and bond angle of 2.794 \AA\ and 122.4$^{0}$, respectively.
In order to determine the structural
phase transition between competing structures, the enthalpy $H$ of each phase
of ReN$_{2}$ is computed.
The enthalpy difference per formula unit (f.u.) as a function of pressure
is plotted in Fig. 1. We do not observe structural phase transition
from $Pbcn$ phase to others up to 100 GPa.
It is interesting to notice that $Pbcn$ structure
 has been observed in experiment being the ground state of ReO$_2$.\cite{li}

VED is defined as the total number of valence electron divided by volume per formula unit,
which is an important factor when searching the superhard materials.
The valence electron shell of Re is 6s$^{2}$ and 5d$^{5}$, and
that of N is 2s$^{2}$p$^{3}$ so that
the total number of valence electrons is 17 for ReN$_2$.
VED obtained in four phases (Table I) is slightly lower than that of Re metal
(0.4761 electrons/\AA$^{3}$) due to the adding of N atoms. High
VED is good to resist the fracture, which
contributes to high bulk pressure of ReN$_2$ (Table I), 
predicting that ReN$_{2}$ could be considered as a candidate of a superhard
material.


\subsection{Elastic property}
The elastic constants of $Pbcn$ structure obtained from our calculations
(\emph{c}$_{11}$=365 GPa, \emph{c}$_{22}$=553 GPa,
\emph{c}$_{33}$=610 GPa, \emph{c}$_{44}$=138 GPa, \emph{c}$_{55}$=230 GPa,
\emph{c}$_{66}$=83 GPa, \emph{c}$_{12}$=307 GPa, \emph{c}$_{13}$=232 GPa,
\emph{c}$_{23}$=242 GPa) satisfy the mechanical
stability requirement well, implying that the orthorhombic
$Pbcn$ structure is mechanically stable. As is concluded
above, ReN$_{2}$ has a very high bulk modulus,
and therefore can be regarded as a candidate of a superhard
material. However, a high bulk modulus by itself does not directly
lead to high hardness. In general, both high bulk modulus and
high shear modulus are indispensable to high hardness.
Therefore, the shear modulus must be taken into account when searching for new superhard materials.
The bulk modulus represents the resistance to fracture, while the
shear modulus represents the resistance to plastic deformation.
Moreover, two other factors are important for technological and
engineering applications: Young's modulus \emph{E} and Poisson's
ratio $\nu$. Young's modulus, defined as the ratio between stress
and strain, is used to provide a measure of stiffness of the
solid. The larger the value of \emph{E}, the stiffer the
material. The Young's modulus and
Posson's ratio for an isotropic material are given
by
$E=\frac{9BG}{3B+G}$, $\nu= \frac{3B-2G}{2(3B+G)}$,
respectively,\cite{Hill} where B and G represent Hill average bulk modulus and
shear modulus. The calculated bulk modulus \emph{B}, shear modulus
\emph{G}, and Young's modulus \emph{E} of
ReN$_{2}$ with $Pbcn$ structure are 334 GPa, 127 GPa, and 337 GPa, respectively,
of which the bulk modulus of ReN$_{2}$ is slightly smaller
than that of Re metal (360 GPa). It is worthy to note that the
bulk modulus obtained from the elastic constants
agrees well with the one obtained through the fit to the 3rd-order
Birch Murnaghan EOS (B$_{0}$), providing a consistent estimation of
the compressibility for ReN$_{2}$.
Possion's ratio $\nu$ reflects the stability of a crystal against
shear. This ratio can formally take values between -1 and 0.5,
which corresponds respectively to the lower limit where the
material does not change its shape, or to the upper limit when the
volume remains unchanged. Poisson's ratio of $Pbcn$ structure
is 0.3319, which is larger than 0.3, indicating that there exists small
volume change during elastic deformation. Also Possion's ratio
provides more information on the characteristics of the bonding
forces. It has been proved that $\nu$=0.25 is the lower limit for
central-force solids and 0.5 is the upper limit, which corresponds
to infinite elastic anisotropy. The high value $\nu$ for ReN$_2$
means that the central interatomic forces are
central.\cite{Ravindran} In order to predict the brittle and
ductile behavior of solids, Pugh introduced the ratio of the bulk
modulus to shear modulus of polycrystalline phases.\cite{pugh} A high (low)
\emph{B}/\emph{G} value is associated with ductility
(brittleness). The critical value which separates ductile and
brittle materials is about 1.75. The significant difference obtained between bulk and shear modulus
indicates the higher ductility of ReN$_{2}$.

Now, we turn to discuss the elastic anisotropy of ReN$_{2}$, which is
estimated by the universal anisotropy index
$A=5\frac{G^{V}}{G^{R}}+\frac{B^{V}}{B^{R}}-6$.\cite{anisotropy}
The discrepancy of $A$ from zero determines the extent of single
crystal anisotropy and accounts for both the shear and the bulk contributions.
The obtained $A$ value is 1.252, indicating that ReN$_2$ has a large elastic anisotropy.
Additionally, 
fractional axis compression as a function of pressure are plotted in Fig. 2.
The curves presented in this figure clearly exhibit that:
a bigger difference in fractional axis compression values
exists for $Pbcn$ phase, implicating that there is a evident structure deformation with increasing pressure;
there is a transition point for compression along $b$- and $c$-axis with increasing pressure, that is,
$b$-axis is more difficult to be compressed than $c$-axis at 38 GPa above.
This abnormality for compression
along different axis can foretell that there is a structural phase transition at about 40 GPa.
Further analysis of the change of the structural parameters
(atomic coordination, bond length, et al.) 
shows that from 40 to 45 GPa fractional coordinations $x$ and $z$ in 8$d$ ($x$, $y$, $z$) sites (N atoms)
change much, yielding the great change of the nearest distance between N atoms ($d_{N-N}$).
In detail, at 40 GPa, $x$ and $z$ are 0.2184 and 0.0752, respectively.
However, $x$ and $z$ reduce to 0.1385 and 0.0349 at 45 GPa, respectively.
Correspondingly, the $d_{N-N}$ reduces from 2.359 \AA\ at 40 GPa to 1.432 \AA\ at 45 GPa.

Further, the Debye temperature $\Theta_{D}$ which
correlates with many physical
properties of materials, such as specific heat, elastic constants,
and melting temperature is given.
The Debye temperature $\Theta_{D}$ can be calculated by the equation
$\Theta_{D}=\frac{h}{k}\left[\frac{3n}{4\pi}\left(\frac{\rho N_A}{M}\right) \right]^{1/3}
[\frac{2}{3}(\frac{\rho}{G})^{3/2}+\frac{1}{3}
(\frac{\rho}{B+4G/3})^{3/2}]^{-1/3}$,\cite{Anderson}
where $h$ is Plank's constant, $k$ is Boltzmann's constant, $N_A$ is Avogadro's number,
$\rho$ is density, $M$ is the molecular weight and $n$ is the number of atoms in the molecule.
The calculated $\Theta_{D}$ value of 735 K in $Pbcn$ is comparative to
the values of 866.4 $K$ in ReB$_{2}$\cite{Hao}, 850 $K$ in ReO$_2$\cite{li}
 and 691 $K$ in OsN$_{2}$.\cite{Wu}
Employing the correlation between the shear modulus and Vickers
hardness reported by Teter \cite{Teter} for a wide variety of
hard materials, the indentation hardness of
ReN$_{2}$ can be estimated to be 
approximately 17.1GPa
, which is comparable to that of
Si$_{3}$N$_{4}$ (21$\pm3$ GPa).\cite{Teter} It is well known
that the hardness of superhard materials should be higher than 40
GPa, indicating that ReN$_{2}$ is not a
superhard material.
Jhi \emph{et al} \cite{Jhi} proposed that the hardness is
determined by elastic constant \emph{c}$_{44}$ rather than
shear modulus. In our case, we note that there is little difference between
the \emph{c}$_{44}$ value of elastic constant and shear modulus. Accordingly,
both Teter estimation and Jhi's viewpoint of
hardness actually achieve a consistent result for ReN$_{2}$.

\subsection{Electronic property}
In order to understand the nature of elastic property, the
electronic density of states is also analyzed. 
The total and partial density of states (DOS) with respect to
Fermi level is shown in Fig. 3. Finite DOS at Fermi level
\emph{N}(E$_{F}$) of 5.035 states/eV shows that ReN$_{2}$ is
metallic. The electrons from Re-5\emph{d} and the N-2\emph{p}
states both contribute much to the DOS near the Fermi level. That is,
the electronic structure of ReN$_{2}$ is governed by strong
hybridization between the Re-5$d$ and N-2$p$ states, imaging that
there is a covalent bond in between Re and N atoms. The existence of covalent
bonding contributes to large bulk modulus, i.e., strong incompressibility.
However, we have not found any pseudogap
that observed in ReB$_{2}$,\cite{Hao,Li}
OsB$_{2}$\cite{Gou} and OsN$_{2}$,\cite{Wu} implying
that covalent bonding in between Re and N atoms is not so strong.
Besides, the calculated net charges on N and Re atom are -0.57 e and 1.15 e,
which indicates that the chemical bonding between Re and N have some characteristics of ionicity.
The occurrence of ionic bonding results in the smaller shear modulus,
which explains the reason of the not so large hardness in ReN$_2$.


\section{Conclusion}
In conclusion, the most stable structure of ReN$_{2}$ is explored based
on first-principles methods in the framework of density
functional theory within local density approximation. It results that orthorhombic $Pbcn$
structure is the most stable structure at ambient pressure.
ReN$_{2}$ is a metallic, super-incompressible solid and presents large elastic anisotropy.
The estimated hardness is comparative to that of Si$_{3}$N$_{4}$.
The structural determination and the discussion of the elastic and electronic properties may be
of use in the synthesis and high pressure study of ReN$_{2}$.

\section{Acknowledgement}
This work was supported by the special Funds for
Major State Basic Research Project of China(973) under grant no.
2007CB925004, 863 Project, Knowledge Innovation Program of Chinese Academy of
Sciences, and Director Grants of CASHIPS. Part of the calculations
were performed in Center for Computational Science of CASHIPS
and the Shanghai Supercomputer Center.

\newpage
\noindent {\bf {\large {TABLE CAPTIONS}}} \vglue 1.0cm \noindent
{\bf {TABLE I:}} Equilibrium lattice
parameters,\emph{V}$_{0}$(\AA$^{3}$), \emph{a} (\AA),
\emph{b}(\AA), \emph{c} (\AA),  density $\rho$ (g/cm$^{3}$),
valence-electron density $\rho_e$ (electrons/\AA$^3$), bulk
modulus \emph{B}$_{0}$ (GPa), and
 enthalpy difference per chemical formula unit (f.u.) $\triangle$H (eV).
 \emph{V}$_{0}$ is of per chemical f.u. 

\newpage

\noindent {\bf {\large {FIGURE CAPTIONS}}}

\vglue 1.0cm

\noindent {\bf {Fig.1:}} Enthalpy difference per formula unit as a
function of pressure. $P4_2/mnm$ is taken as a reference point.

\vglue 1.0cm

\noindent {\bf {Fig.2:}} Fractional axis compression
as a function of pressure for $Pbcn$ structure.

\vglue 1.0cm

\noindent {\bf {Fig.3:}} Total and partial density of states for $Pbcn$ structure. Vertical dotted
line indicates the Fermi level.

\newpage

\begin{table}[htbp]
\begin{center}
\caption{\hspace{12pt}Li \textit {et al.}} \vspace{0.1cm}
\begin{tabular}{cccccc}
\hline \hline
           &\emph{Pbcn}    &$P4_2/mnm$     &$Fm\bar{3}m$  &$Pmmn$ &Reference \\
\hline
\emph{V}$_{0}$    &31.251        &31.693      &27.712  &26.779 &This work    \\
                  &               &31.7         &27.7     &26.7 &[10]  \\
\emph{a}          &4.518         &4.769       &4.801   &8.088  &This work     \\
                  &               &4.775        &4.803    &8.083 &[10] \\
\emph{b}          &5.657         &             &         &2.380  &This work     \\
                  &               &             &         &2.385 &[10]  \\
\emph{c}          &4.902         &2.789       &         &2.782  &This work     \\
                  &               &2.78         &         &2.77 &[10]   \\
$\rho$            &11.385        &11.217      &12.838  &12.406 &This work \\
$\rho_{e}$        &0.416         &0.410       &0.469   &0.453  &This work \\
B$_{0}$           &331.3          &337.8        &381.9    &365.2     &This work \\
$\triangle$H       &0              &0.015        &1.457    &1.663     &This work \\

\hline \hline
\end{tabular}
\end{center}
\end{table}

\clearpage
\newpage


\begin{figure}[htbp]
 \includegraphics[width=3.0 in, angle=0,clip]{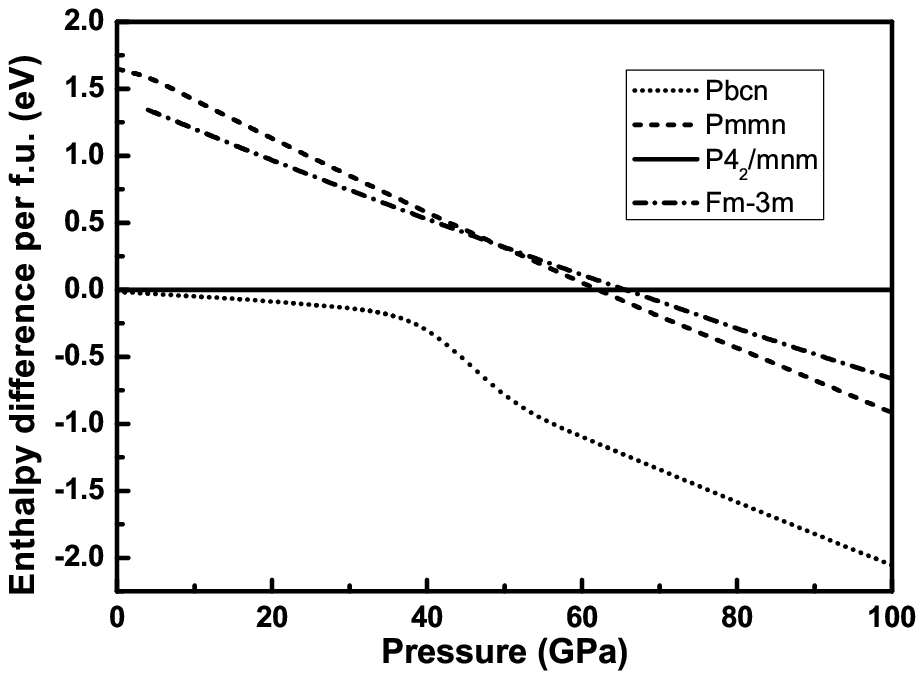}
 \caption{\hspace{12pt} Li \textit {et al.}}
\end{figure}

\clearpage
\newpage
\begin{figure}[htbp]
 \includegraphics[width=8.0 cm, angle=0,clip]{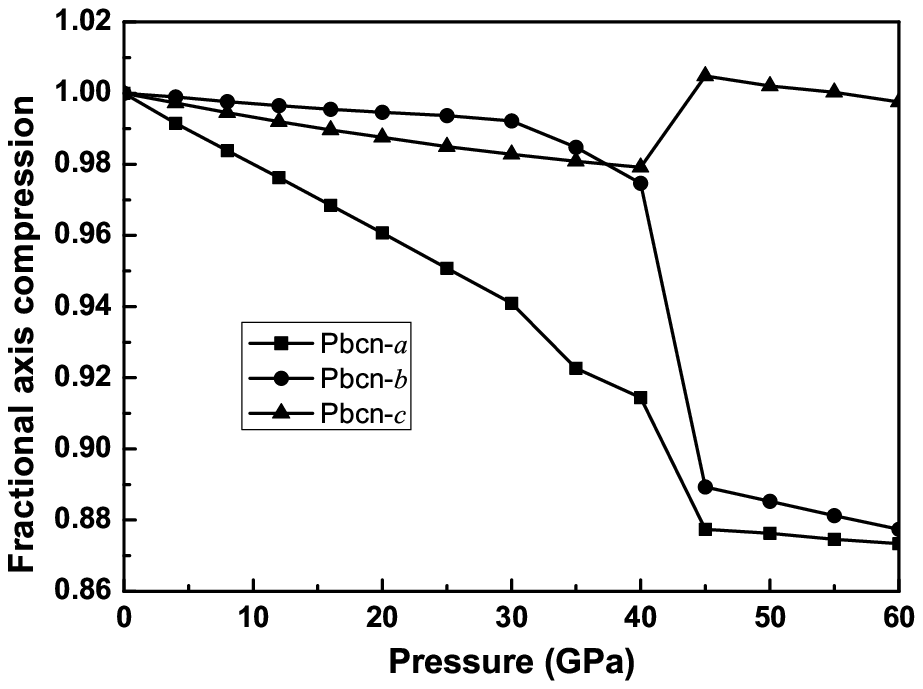}
 \caption{\hspace{12pt} Li \textit {et al.}}
\end{figure}

\clearpage
\newpage
\begin{figure}[htbp]
\includegraphics[width=8.0 cm, angle=0,clip]{fig3.eps}
 \caption{\hspace{12pt} Li \textit {et al.}}
\end{figure}


\begin{thebibliography}{00}
\bibitem{Crowhurst}
J. C. Crowhurst, A. F. Goncharov, B. Sadigh, C. L. Evans, P. G.
Morrall, J. L. Ferreira, and A. G. Nelson, Science 311 (2006) 1275.

\bibitem{Kaner}
R. B. Kaner, J. J. Gilman and S. H. Tolbert. Science 308 (2006)
1268.

\bibitem{Chung}
H-y. Chung, M. B. Weinberger, J. B. Levine, A. Kavner, J-M. Yang,
S. H. Tolbert, and R. B. Kaner, Science 316 (2007) 436.

\bibitem{Young}
A. F. Young, C. Sanloup, E. Gregoryanz, S. Scandolo, R. J. Hemley,
and H. K. Mao, Phys. Rev. Lett. 96 (2006) 155501.

\bibitem{Young-1}
A. F. Young, J. A. Montoya, C. Sanloup, M. Lazzeri, E. Gregoryanz,
and S. Scandolo, Phys. Rev. B 73 (2006) 153102.

\bibitem{Wu}
Z. Wu, X. Hao, X. Liu, and J. Meng, Phys. Rev. B 75 (2007) 054115.

\bibitem{Chen}
Z. W. Chen, X. J. Guo, Z. Y. Liu, M. Z. Ma, Q. Jing, G. Li, X. Y.
Zhang, L. X. Li, Q. Wang, Y. J. Tian, and R. P. Liu, Phys. Rev. B
75 (2007) 054103.

\bibitem{Yu}
 R. Yu, Q. Zhan, and L-C. De Jonghe, Angew. Chem. Int. Ed. 46 (2007) 1136.

\bibitem{Placa}
S. La Placa and B. Post, Acta Crystallogr. 15 (1962) 97.

\bibitem{cms}
E. Zhao and Z. Wu, Comput. Mat. Sci. 44 (2008) 531.

\bibitem{Segall}
M. D. Segall, P. J. D. Lindan, M. J. Probert, C. J. Pickard, P. J.
Hasnip, S. J. Clark, M. C. Payne, J. Phys.: Cond. Matt. 14 (2002)
2717.

\bibitem{LDA}
D. M. Ceperley and B. J. Alder, Phys. Rev. Lett. 45 566 (1980); J.
P. Perdew and Y. Wang, Phys. Rev. B 45 (1992) 13244 .

\bibitem{BFGS}
B. G. Pfrommer, M. Cote, S. G. Louie, and M. L. Cohen, J. Comput.
Phys. 131 (1997) 133.

\bibitem{Ultrasoft}
D. Vanderbilt, Phys. Rev. B 41 (1990) R7892.

\bibitem{Hill}
R. Hill, Proc. Phys. Soc. London  65 (1952) 349.

\bibitem{Beckstein}
O. Beckstein, J. E. Klepeis, G. L. W. Hart, and O. Pankratov,
Phys. Rev. B 63 (2001) 134112.

\bibitem{Born}
M. Born and K. Huang, Dynamical Theory of Crystal
Lattices, Clarebdom, Oxford, 1956.


\bibitem{3rdEOS}
J. -P. Poirier, Introduction to the physics of the Earth's
Interior, Cambridge University Press, Cambridge, 2000.

\bibitem{li}
Y. L. Li and Z. Zeng, Chin. Phys. Lett 25 (2008) 4086.

\bibitem{Ravindran}
P. Ravindran, L. Fast, P. A. Korzhavyi, B. Johansson, J. Wills,
and O. Eriksson, J. Appl. Phys. 84 (1998) 4891.

\bibitem{pugh}
S. F. Pugh, Philos. Mag. 45 (1954) 823.

\bibitem{anisotropy}
S. I. Ranganathan and M. Ostoja-Starzewski, Phys. Rev. Lett. 101 (2008) 055504 .

\bibitem{Anderson}
O. L. Anderson, J. Phys. Chem. Solids 24 (1963) 909.


\bibitem{Hao}
X. Hao, Y. Xu, Z. Wu, D. Zhou, X. Liu, X. Cao, and J. Meng, Phys.
Rev. B 74 (2006) 224112.

\bibitem{Teter}
D. M. Teter, MRS. Bull. 23 (1998) 22.

\bibitem{Jhi}
S-H. Jhi, J. Ihm, S. G. Louie, and M. L. Cohen, Nature 399 (1999)
132.

\bibitem{Li}
Y. L. Li, G. H. Zhong, and Z. Zeng, accepted by Chin. Phys. B (2009)

\bibitem{Gou}
H. Y. Gou, L. Hou, J. W. Zhang, H. Li, G. F. Sun, and F. M. Gao,
Appl. Phys. Lett. 88 (2006) 221904.

\end{thebibliography}
\end{document}